\documentclass[fleqn,10pt,onecolumn]{wlscirep}
\usepackage{amsfonts}
\usepackage{txfonts}
\usepackage{amsmath}
\usepackage{float}
\usepackage{graphicx}
\usepackage{dcolumn}
\usepackage{bm}
\usepackage{amssymb}
\usepackage{mathrsfs}
\usepackage{amsmath}
\usepackage{bm}
\usepackage{subfigure}
\usepackage{cases}
\newcommand{\bra}[1]{\left\langle #1\right|}
\newcommand{\ket}[1]{\left|#1\right\rangle}

\title{Multi-qubit Quantum Rabi Model and Multi-partite Entangled States in a Circuit QED System}
\author[1]{Jialun Li}
\author[1,*]{Gangcheng Wang}
\author[1]{Ruoqi Xiao}
\author[1,$\dag$]{Chunfang Sun}
\author[2]{Chunfeng Wu}
\author[1,$\ddag$]{Kang Xue}
\affil[1]{Center for Quantum Sciences and School of Physics, Northeast Normal University, Changchun 130024, China}
\affil[2]{Science and Mathematics, and Pillar of Engineering Product Development, Singapore University of Technology and Design, 8 Somapah Road, Singapore 487372}
\affil[*]{Gangcheng Wang (Email: wanggc000@163.com)}
\affil[$\dag$]{Chunfang Sun (Email: suncf997@nenu.edu.cn)}
\affil[$\ddag$]{Kang Xue (Email: xuekang@nenu.edu.cn)}

\begin{abstract}
Multi-qubit quantum Rabi model, which is a fundamental model describing light-matter interaction, plays an important role in various physical systems. In this paper, we propose a theoretical method to simulate multi-qubit quantum Rabi model in a circuit quantum electrodynamics system. By means of external transversal and longitudinal driving fields, an effective Hamiltonian describing the multi-qubit quantum Rabi model is derived. The effective frequency of the resonator and the effective splitting of the qubits depend on the external driving fields. By adjusting the frequencies and the amplitudes of the driving fields, the stronger coupling regimes could be reached. The numerical simulation shows that our proposal works well in a wide range of parameter space. Moreover, our scheme can be utilized to generate two-qubit gate, Schr\"{o}dinger states, and multi-qubit GHZ states. The maximum displacement of the Schr\"{o}dinger cat states can be enhanced by increasing the number of the qubits and the relative coupling strength. It should be mention that we can obtain high fidelity Schr\"{o}dinger cat states and multi-qubit GHZ states even the system suffering dissipation. The presented proposal may open a way to study the stronger coupling regimes whose coupling strength is far away from ultrastrong coupling regimes.
\end{abstract}

\begin{document}
\flushbottom
\maketitle
\thispagestyle{empty}

\section*{Introduction}
The quantum Rabi model (QRM) \cite{Rabi1936,Rabi1937,braak2011}, which is a fundamental model to describe the interactions between light and matter, occupies a crucial position in various physical systems, such as quantum optics \cite{scully}, solid state system \cite{irish2007}, molecular system \cite{thanopulos2004}, and so on. When the ratio between the coupling strength ($g$) and the mode frequency ($\omega$) satisfies $g/\omega \ll 1$, the rotating-wave approximation (RWA) is suitable, and the counter-rotating term (CRT) can be ignored. In this case, the QRM is reduced to the Jaynes-Cummings (JC) model \cite{Jaynes1963}, which has been applied to explain many physical phenomena, such as the revivals of the atomic population inversion after its collapse \cite{Cummings1965,Eberly1980}, vacuum Rabi splitting \cite{Thompson1992,Boca2004}, and so on. Recently, new coupling regimes, such as ultra-strong coupling (USC) and deep-strong coupling (DSC) regimes, have been reached in some circuit QED systems \cite{Michel2007,Bourassa2009,Casanova2010,Wallraff2004,Srinivasan2011,Niemczyk2010}. In this case, the CRT cannot be neglected. Consequently, many interesting effects induced by CRT appear in these regimes \cite{garz2016,wangx2017,Ridolfo2012,Ridolfo2013,Law2013,caox2010,aiq2010,lipb2012,wangx2014,reiter2013,hes2014,Huang2017}. The implementations of QRM in USC and DSC regimes have also motivated new applications to the quantum information processing \cite{rossatto2016,felicettiprl2014,kyawprb2015,romero2012,wangym2017,cui2018}. It should be mentioned that, though great progress has been achieved, it is also challenging to implement QRM in USC, and DSC regimes experimentally.

Of particular interest is how to simulate the QRM in USC, and even DSC regimes when the system is far from USC regime. Motivated by this consideration, some quantum simulation approaches have been proposed in various physical systems, such as superconducting circuits \cite{Deng2015,Ballester2012,Li2013,Braumuller2017,wangym2018}, quantum optical \cite{Crespi2012}, trapped ions \cite{Pedernales2015,Aedo2018}, cold atoms \cite{Felicetti2017,Felicetti2017_2,Schneeweiss2018}, and so on. These quantum simulation proposals provide us with experimental feasible methods to implement QRM in USC and DSC regimes. Very recently, the quantum simulations of USC and DSC regimes extend to the multi-qubit case with trapped ions and anisotropic quantum Rabi model with superconducting circuits. The simulations of the generalized models provide us with platforms to study concerning physical issues, such as quantum critical phenomena, multi-partite entanglement, and so on.

On the other hand, the qubit-dependent displacement interaction describes a quantum resonator conditionally displaced according to qubit(s)' states. Such interaction plays an important role in understanding the fundamentals of quantum physics \cite{Leggett2002,Armour2002,Liao2016,Haljan2005,Yin2013,Liu2005,Liao2008}. Based on such type interaction, the superposition of the coherent states can be prepared in various of systems \cite{Armour2002,Liao2016,Haljan2005,Yin2013,Liu2005,Liao2008}. The qubit-dependent displacement interaction also has been used to the quantum information processing \cite{Sorensen2000,Zoller2003,Leibfried2003,Feng2007,Feng2009,Billangeon2015,Zhu2003,Kirchmair2009,Wang2010,MS1999,Wang2002-1,Christian2008}, such as generation of unconventional phase gate \cite{Zhu2003} and multipartite entangled states \cite{Kirchmair2009,Wang2010,MS1999}. Following the theoretical and experimental study of the QRM\cite{Li2013}, we focus on the simulation of multi-qubit QRM, and we will study its applications to generation of two-qubit quantum gate, Schr\"{o}dinger cat states and multi-qubit GHZ states.

In this paper, we propose an alternative scheme to simulate multi-qubit QRM in USC regime, and even DSC regime with a circuit QED setup. The system consists of multiple flux qubits, which strongly coupled to a resonator. To obtain the tunable multi-qubit QRM, we apply transversal and longitudinal external driving fields on the qubits. We show the stronger coupling regimes can be reached by tuning the driving amplitudes and frequencies. Additionally, we study some applications of simulated Hamiltonian on two-qubit quantum gate, superposition coherent states and multi-qubit entangled states. The results show that the non-trivial two-qubit gate is equivalent to the controlled-NOT (CNOT) gate. Based on the multi-qubit conditional interaction Hamiltonian, the Schr\"{o}dinger cat states and multi-qubit GHZ states can be generated. The maximum displacement of the Schr\"{o}dinger cat states depends on the number of qubits and the relative coupling strength, which indicates the maximum displacement can be enhanced by increasing number of the qubits and the relative coupling strength.

\section*{The derivation of the effective Hamiltonian}
In this section, we first derive a effective QRM, in which the relative coupling strength can be adjusted by tuning the frequency of the external driving fields. We also show the fidelity of the simulated Hamiltonian. We consider $N$ qubits strongly coupled to a single-mode harmonic oscillator. The qubits are driven by the longitudinal and transversal external driving fields. Such model can be realized in a variety of physical systems. Here we adopt a circuit QED setup to demonstrate our proposal. We consider $N$ flux qubits are coupled to a transmission line resonator, which can be modeled as a single mode harmonic oscillator. Assuming the qubits are tuned to the degeneracy point, then the Hamiltonian in this case reads (here and after, we set $\hbar=1$)
\begin{equation}\label{eq_rabi_1}
\hat{H}=\hat{H}_{0}+\hat{H}_{\rm int}+\hat{H}_{\rm d},
\end{equation}
where
\begin{subequations}
\label{eq_rabi_2}
\begin{align}
\hat{H}_{0}&=\omega_{r}\hat{a}^{\dag}\hat{a} + \frac{1}{2}\sum_{k=1}^{N}\varepsilon_{k} \hat{\sigma}_{k}^{z}, \\
\hat{H}_{\rm int}&=\frac{1}{2}\sum_{k=1}^{N}g_{k}(\hat{a}^{\dag}+\hat{a})\hat{\sigma}_{k}^{x},  \\
\hat{H}_{\rm d}&=\frac{1}{2}\sum_{k=1}^{N}\Omega_{z}\cos(\omega_{z} t)\hat{\sigma}_{k}^{z} + \frac{1}{2}\sum_{k=1}^{N}\Omega_{x}\cos(\omega_{x} t)\hat{\sigma}_{k}^{x}.
\end{align}
\end{subequations}
 Here the operator $\hat{a}$ ($\hat{a}^{\dag}$) is the annihilation (creation) operator of the bosonic field with frequency $\omega_{r}$. The qubits are described by Pauli matrices $\hat{\sigma}^{\alpha}_{k}$ ($\alpha=x,y,z$), which denotes $\alpha$ component of the $k$-th Pauli matrix. For simplicity, we consider all the qubits possess the same energy splitting $\varepsilon$ (i.e., $\varepsilon_{k}=\varepsilon$), and the qubits couple to the bosonic field with unified coupling strength $g$ (i.e. $g_{k}=g$). $\hat{H}_{\rm int}$ shows the interaction between the resonator and the qubits. All the qubits are driven by two classical fields with the frequencies $\omega_{z}$ and $\omega_{x}$, and the corresponding amplitudes are denoted by $\Omega_{z}$ and $\Omega_{x}$. In this case, we introduce the collective operators $\hat{J}_{\alpha}=\frac{1}{2}\sum_{k=1}^{N}\hat{\sigma}_{\alpha}$ to simplify the Hamiltonian (\ref{eq_rabi_1}) as $\hat{H}_{0}=\omega_{r}\hat{a}^{\dag}\hat{a}+\varepsilon \hat{J}_{z}$, $\hat{H}_{\rm int}=g(\hat{a}^{\dag}+\hat{a})\hat{J}_{x}$, and $\hat{H}_{\rm d}=\Omega_{z}\cos(\omega_{z}t)\hat{J}_{z}+\Omega_{x}\cos(\omega_{x}t)\hat{J}_{x}$. Choosing the rotating framework defined by
 \begin{equation}\label{eq_ut}
   U(t)=\exp\left(-i\frac{\Omega_{x}}{2}\hat{J}_{x}t\right)\exp\left(-i\omega_{x}\hat{J}_{z}t-i\omega_{x}\hat{a}^{\dag}\hat{a}t\right)
\end{equation}
and considering the following conditions
\begin{equation}
\label{eq_conditions}
\Omega_{x}=2\omega_{z},\quad \omega_{x}\gg \Omega_{x} \gg g, \quad \omega_{z}\gg\Omega_{z},\quad \omega_{z}\gg\varepsilon-\omega_{x},
\end{equation}
we can neglect the fast oscillating terms and obtain the following time-independent effective Hamiltonian (the detailed derivation is shown in the {\bf Methods} section)
\begin{equation}
\label{eq_rabi_eff}
  \hat{H}_{\rm eff}=\tilde{\omega}_{r}\hat{a}^{\dag}\hat{a}+\tilde{\varepsilon}\hat{J}_{z}+
\tilde{g}(\hat{a}+\hat{a}^{\dag})\hat{J}_{x},
\end{equation}
where $\tilde{\omega}_{r}=\omega_{r}-\omega_{x}$, $\tilde{\varepsilon}=\Omega_{z}/2$, and $\tilde{g}=g/2$ are the effective energy splitting of the qubits, effective frequency of the resonator, and effective coupling strength, respectively. Such effective Hamiltonian describes a multi-qubit generalization of quantum Rabi model (i.e., Dicke model), in which the frequency of the resonator and the energy splitting of qubits can be adjusted by tuning the frequencies and amplitudes of external driving fields. The relative coupling strength reads
\begin{equation}
\label{eq_geff}
\frac{\tilde{g}}{\tilde{\omega}_{r}}=\frac{g}{2(\omega_{r}-\omega_{x})},
\end{equation}
The relative coupling strength can be adjusted by tuning the frequency of the transversal driving fields. Thus we can obtain the multi-qubit QRM in different coupling regimes.
 \begin{figure*}[t]
  \centering
  \includegraphics[width=0.8\textwidth]{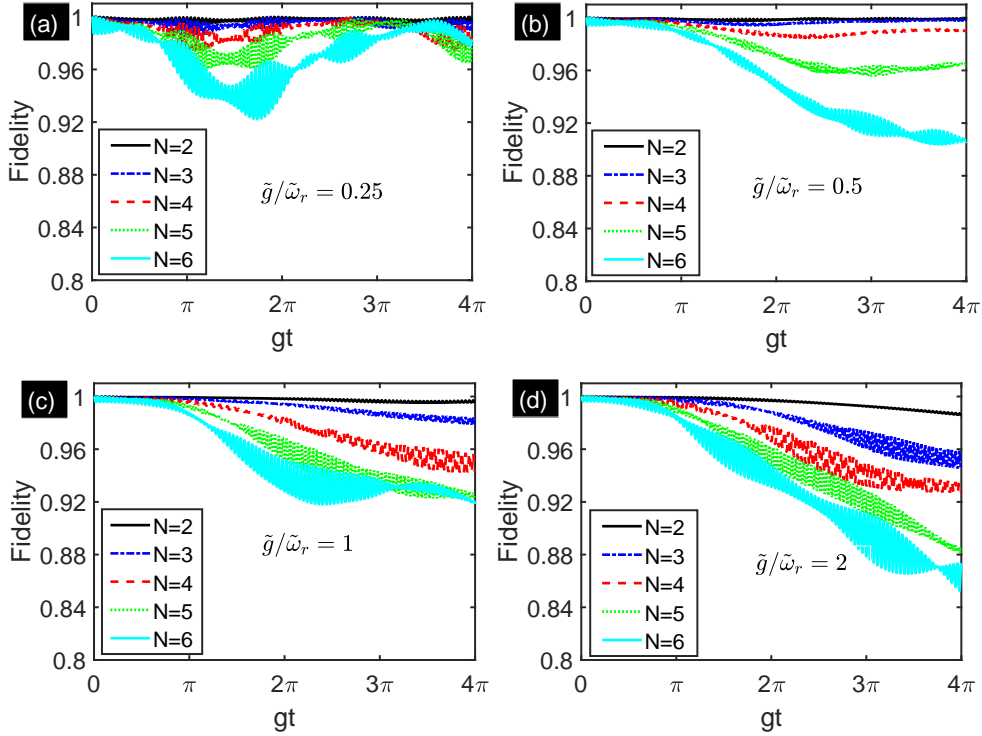}\\
    \caption{The fidelity of the evolution states as a function of evolution time for multi-qubit under different relative coupling strength. (a) the fidelity of the evolution states under the relative coupling strength $\tilde{g}/\tilde{\omega}_{r}=0.25$. (b) the fidelity of the evolution states under the relative coupling strength $\tilde{g}/\tilde{\omega}_{r}=0.5$. (c) the fidelity of the evolution states under the relative coupling strength $\tilde{g}/\tilde{\omega}_{r}=1$. (d) the fidelity of the evolution states under the relative coupling strength $\tilde{g}/\tilde{\omega}_{r}=2$. The frequencies of the transversal driving are $\omega_{x}=\left\{0.996,0.998,0.999,0.9995\right\}\omega_{r}$ for (a), (b), (c) and (d), respectively. The other parameters are $\varepsilon=\omega_{r}$, $\Omega_{z}=0.004\omega_{r}$, $\Omega_{x}=2\omega_{z}=0.2\omega_{r}$, and $g=0.002\omega_{r}$. We choose $\ket{\psi(0)}=\ket{gg\cdots g}$ as initial state.}
  \label{fig1}
\end{figure*}

 In order to assess the validity of the effective Hamiltonian. We compare the time-dependent evolution states governed by the full Hamiltonian~(\ref{eq_rabi_1}) and the effective Hamiltonian~(\ref{eq_rabi_eff}). Let $\ket{\psi(0)}=\ket{gg\cdots g}\otimes \ket{0}_{r}$ be the initial state and the evolution states governed by the Hamiltonian~(\ref{eq_rabi_1}) and~(\ref{eq_rabi_eff}) are denoted by $\ket{\psi(t)}$ and $\ket{\tilde{\psi}(t)}_{\rm ideal}$, respectively. We denote the evolution state governed by Hamiltonian (\ref{eq_rabi_1}) in the rotating framework defined by $U(t)$ with $\ket{\tilde{\psi}(t)}=U^{\dag}(t)\ket{\psi(t)}$. The fidelity of the evolution states $\ket{\tilde{\psi}(t)}$ and $\ket{\tilde{\psi}(t)}_{\rm ideal}$ reads ${\rm F}(t)=\left|\left\langle\tilde{\psi}(t)|\tilde{\psi}(t)\right\rangle_{\rm ideal}\right|^{2}$. Considering the approximate conditions, we choose the following parameters: $\varepsilon=\omega_{r}$, $\Omega_{z}=0.004\omega_{r}$, $\Omega_{x}=2\omega_{z}=0.2\omega_{r}$, $g=0.002\omega_{r}$ and $\omega_{x}=\left\{0.996,0.998,0.999,0.9995\right\} \omega_{r}$. Under such parameters, the relative coupling strength are $\tilde{g}/\tilde{\omega}_{r}=\left\{0.25,0.5,1,2\right\}$ and the system is driven to stronger coupling regimes. In Fig.~\ref{fig1}, we plot the fidelity of evolution states for $N=2$ (black solid line), $N=3$ (blue dash-dotted line), $N=4$ (red dashed line), $N=5$ (green dotted line) and $N=6$ (cyan solid line). The Fig.~\ref{fig1}(a)-\ref{fig1}(d) show the fidelity when the relative coupling strength $\tilde{g}/\tilde{\omega}_{r}=\left\{0.25,0.5,1,2\right\}$, respectively. The results show that the effective Hamiltonian is validity when the number of the qubits and the relative coupling strength are not very large.

\section*{The applications of the effective Hamiltonian}
In this section, we will illustrate some applications to the simulated multi-qubit QRM on quantum information processing. Such as the generation of quantum gate, the Schr\"{o}dinger cat states, and multi-qubit GHZ states. Moving to the rotating frame associated with $U'_{3}=\exp\left(-i\tilde{\varepsilon}\hat{J}_{z}t-i\tilde{\omega}_{r}\hat{a}^{\dag}\hat{a}t\right)$, the effective Hamiltonian is recast as following form
\begin{equation}\label{eq_rabi_7}
  \hat{H}^{I}_{\rm eff}(t)=\frac{\tilde{g}}{2}\left(\hat{J}_{+}e^{i\tilde{\varepsilon}t} + \hat{J}_{-}e^{-i\tilde{\varepsilon}t}\right)\left(\hat{a}e^{-i\tilde{\omega}_{r}t}+\hat{a}^{\dag}e^{i\tilde{\omega}_{r}t}\right).
\end{equation}
If we consider all the qubits have zero effective energy splitting (i.e., $\tilde{\varepsilon} = 0$), the Eq.~(\ref{eq_rabi_7}) can be reduced to the following form
\begin{equation}\label{eq_rabi_7-1}
\hat{H}^{I}_{\rm eff}(t)=\tilde{g}\left(\hat{a}e^{-i\tilde{\omega}_{r}t}+\hat{a}^{\dag}e^{i\tilde{\omega}_{r} t}\right)\hat{J}_{x}.
\end{equation}
This is a periodic Hamiltonian with period $T=2\pi/|\tilde{\omega}_{r}|$. The evolution operator for Hamiltonian (\ref{eq_rabi_7-1}) can be obtained by means of the Magnus expansion \cite{Blanes2009}
\begin{equation}
\label{eq_rabi_9}
\mathcal{U}(t)=\exp\left(\Omega_{1}(t)+\Omega_{2}(t)\right),
\end{equation}
where
 \begin{subequations}
 \label{eq_rabi_10}
\begin{align}
\Omega_{1}(t)&=\frac{\tilde{g}}{\tilde{\omega}_{r}}\left[\hat{a}^{\dag}(1- e^{i\tilde{\omega}_{r}t})-\hat{a}(1-e^{-i\tilde{\omega}_{r}t})\right]\hat{J}_{x},\\
\Omega_{2}(t)&=i\frac{\tilde{g}^{2}}{\tilde{\omega}_{r}^{2}}\left(\tilde{\omega}_{r}t-\sin\tilde{\omega}_{r}t\right)\hat{J}_{x}^{2}.
\end{align}
\end{subequations}
Considering the commutator $[\Omega_{1}(t),\Omega_{2}(t)]=0$, the evolution operator can be recast as follows
 \begin{equation}
 \label{eq_rabi_11}
  \mathcal{U}(t)=D\left(\beta(t)\hat{J}_{x}\right)\exp\left(i\phi(t)\hat{J}_{x}^{2}\right),
 \end{equation}
where the displacement operator is given by $D(\beta)=\exp\left(\beta \hat{a}^{\dag}-\beta^{*}\hat{a}\right)$. The parameters $\beta(t)$ and $\phi(t)$ are defined as $\beta(t)=(\tilde{g}/\tilde{\omega}_{r})(1-e^{i\tilde{\omega}_{r}t})$ and $\phi(t)=(\tilde{g}/\tilde{\omega_{r}})^{2}(\tilde{\omega}_{r}t-\sin\tilde{\omega}_{r}t)$. For the following convenience, we introduce the collective states, which is the eigenstates of the collective operators $\left\{\hat{J}^{2},\hat{J}_{\alpha}\right\}$. Let the collective states $\left\{\ket{j,j_{\alpha}}_{\alpha};~j_{\alpha}=-j,-j+1,\cdots j;~\alpha=x,y,z\right\}$ be the eigenstates of operator set $\{\hat{J}^{2},\hat{J}_{\alpha}\}$, and they satisfy the following equations: $\hat{J}^{2}\ket{j,j_{\alpha}}_{\alpha}=j(j+1)\ket{j,j_{\alpha}}_{\alpha}$, $\hat{J}_{\alpha}\ket{j,j_{\alpha}}_{\alpha}=j_{\alpha}\ket{j,j_{\alpha}}_{\alpha}$.

In the following, we will use the evolution operator given in Eq.~(\ref{eq_rabi_11}) to generate two-qubit quantum gate, Schr\"{o}dinger cat state, and $N-$qubit GHZ states. To describe the dynamics of the system under dissipation, we utilize the following master equation
\begin{equation}\label{eq_master_x}
  \dot{\rho}=-i\left[\hat{H}(t),\rho\right]+\gamma\sum_{k=1}^{N}\mathcal{L}[\hat{\sigma}_{k}^{-}]\rho(t)+\kappa\mathcal{L}[\hat{a}]\rho(t),
\end{equation}
where $\rho(t)$ is the time-dependent density matrix. The time-dependent density matrix in the rotating framework can be obtained by $\tilde{\rho}=U(t)\rho U^{\dag}(t)$ and its dynamics is governed by the Hamiltonian $\hat{\tilde{H}}(t)=U^{\dag}H(t)U(t)-iU^{\dag}(t)\partial_{t}U(t)$, which is full Hamiltonian in the rotating framework. The qubits decay rate and resonator loss rate are denoted by $\gamma$ and $\kappa$, respectively. $\mathcal{L}[\hat{A}]\rho=\frac{1}{2}\left(2\hat{A}\rho \hat{A}^{\dag}-\hat{A}^{\dag}\hat{A}\rho-\rho \hat{A}^{\dag}\hat{A}\right)$ is the Lindblad superoperator describing the losses of the system. In the following numerical simulation, we adopt the following realistic parameters \cite{Liu2007,Goerz2017}: $\varepsilon=\omega_{r}=2\pi\times 10$ GHz, $\Omega_{x}=2\omega_{z}=2\pi\times 2$ GHz, $g=2\pi\times 20$ MHz and $\omega_{x}=2\pi\times 9.98$ GHz. The decay rate of the qibit and resonator loss rate are taken as $\gamma=2\pi\times 0.05$ MHz and $\kappa = 2\pi\times 0.012$ MHz. We switch off the longitudinal driving fields (i.e., $\Omega_{z}=0$). The parameters are list in Table (\ref{tab1}). Under such parameters, the relative coupling strength is $\tilde{g}/\tilde{\omega}_{r}=0.5$ and the effective energy splitting is $\tilde{\varepsilon}=0$.
\begin{table}
\centering
\caption{The realistic parameters in the numerical simulation are listed in the following table \cite{Liu2007,Goerz2017}.}
\label{tab1}
\begin{tabular}{ccccccc}
\hline
\hline
$\varepsilon/2\pi$ & $\omega_{r}/2\pi$ & $g/2\pi$ & $\Omega_{x}/2\pi$ & $\omega_{x}/2\pi$ & $\gamma/2\pi$ & $\kappa/2\pi$\\
\hline
10~GHz & 10~GHz & 20~MHz & 2~GHz & 9.98~GHz & 0.05~MHz & 0.012~MHz\\
\hline
\hline
\end{tabular}
\end{table}
\subsection*{The realization of the quantum gate.}
To obtain the two-qubit quantum gate, we consider $N=2$ and evolution time $t=T=2\pi/\tilde{\omega}_{r}$. In this case, $\hat{J}_{x}=(\hat{\sigma}_{1}^{x}+\hat{\sigma}_{2}^{x})/2$ and the evolution operator (\ref{eq_rabi_11}) reduces to $\mathcal{U}(T)=\cos(\phi/2)\mathcal{I}+\sin(\phi/2)\hat{\sigma}_{1}^{x}\hat{\sigma}_{2}^{x}$ with $\phi=2\pi(\tilde{g}/\tilde{\omega_{r}})^{2}$, where $\mathcal{I}$ is the identity operator for two-qubit system. Here, we have omitted a global phase. Obviously, such quantum gate is capable to generate entanglement when $\phi\neq m\pi$ ($m$ is an integer). To describe the entanglement generation capacity of the unitary operator, we utilize the entangling power given by Zanardi {\it et.al.} \cite{Zanardi2000,Zanardi2001,Wang2002-2,Ma2007}. The entangling power defined on $d\times d$ system can be expressed in terms of the linear entropy of operators $\mathcal{U}$, $\mathcal{U}S_{12}$, and $S_{12}$ as follows
\begin{equation}
\label{eq_rabi_14}
e_{p}(\mathcal{U})=\left(\frac{d}{d+1}\right)^{2}\left[E(\mathcal{U})+E(\mathcal{U}S_{12})-E(S_{12})\right],
\end{equation}
where $S_{12}=\sum_{i,j}|ij\rangle\langle ji|$ is the swapping operator acting on the tensor product space and the linear entropy of the $\mathcal{U}$ is given by $E(\mathcal{U})=1-\frac{1}{d^{4}}Tr[\mathcal{U}^{R}(\mathcal{U}^{R})^{\dag}\mathcal{U}^{R}(\mathcal{U}^{R})^{\dag}]$. The rearrangement of $\mathcal{U}$ is defined as $\mathcal{U}^{R}_{ij,kl}=\mathcal{U}_{ik,jl}$ \cite{Ma2007}. The entangling power of the quantum gate can be obtained as $e_{p}(\mathcal{U})=\frac{2}{9}\sin^{2}\phi$. When $\phi=\pi/2$ (i.e. $\tilde{g}/\tilde{\omega_{r}}=0.5$), we obtain a quantum gate with maximum quantum entangling power. Such non-trivial quantum gate is local equivalent to the CNOT gate \cite{Makhlin2002,Zhang2003}. We can check the following local equivalence relation
\begin{equation}
\label{eq_rabi_18}
{\rm CNOT} = (u_{1}\otimes u_{2})\mathcal{U}(u_{3}\otimes u_{4}),
\end{equation}
where the local unitary operators are as follows
\begin{equation}
\label{eq_rabi_19}
 \begin{array}{llll}
  u_{1}=\frac{1}{\sqrt{2}}\left(
  \begin{array}{cc}
     1 & -1 \\
    -1 & -1 \\
  \end{array}\right), & u_{2}=\left(
                  \begin{array}{cc}
                  1 & 0 \\
                  0 & 1 \\
                  \end{array}
                   \right),&
  u_{3}= \frac{1}{\sqrt{2}}\left(
                   \begin{array}{cc}
                     1 & i \\
                     -1 & i\\
                   \end{array}\right), & u_{4}=\frac{1}{\sqrt{2}}\left(
                                  \begin{array}{cc}
                                    1 & i \\
                                    i & 1 \\
                                  \end{array}
                                \right).
\end{array}
\end{equation}
In order to assess the performance of our proposal to generate CNOT equivalent gate against sources of error, we adopt the process fidelity ${\rm F}_{\rm pro}$, which measures the difference between ideal and real quantum processes. For an ideal unitary process $\mathcal{U}$ and its real process $\mathcal{E}(\mathcal{U})$, the process fidelity reads
\begin{equation}
\label{eq_pro_fid}
\begin{split}
{\rm F}_{\rm pro}(\mathcal{E},\mathcal{U})=\frac{1}{d^{3}}\sum_{j=1}^{d^{2}}{\rm Tr}\left[\mathcal{U}W_{j}^{\dag}\mathcal{U}^{\dag}\mathcal{E}\left(W_{j}\right)\right]
\end{split}.
\end{equation}
For two-qubit system, $d=4$ and $W_{j}$ is the operator basis acting on the 4-dimensional Hilbert space. The operator basis can be represented with the Pauli matrices (i.e., $W_{j}\in \left\{\mathcal{I},\hat{\sigma}_{1}^{x},\cdots, \hat{\sigma}_{1}^{z}\hat{\sigma}_{2}^{z}\right\}$). If we adopt the full Hamiltonian without dissipation (i.e., Eq.~(\ref{eq_rabi_1})) under the parameters listed in Table (\ref{tab1}), the process fidelity can reach $99.57\%$. If we adopt the full Hamiltonian with dissipation (i.e., Eq.~(\ref{eq_master_x})), the process fidelity of the quantum gate is $96.32\%$. The higher performance of the quantum gate needs to resort to adopt superconducting qubit with lower decay rate.

\subsection*{The generation of Schr\"{o}dinger cat states.}
The conditional interaction is also crucial in creating superposed coherent states and hence exploring the superposition rule. We then investigate the creation of the superposed coherent states with the derived effective Hamiltonian. We choose $|\psi(0)\rangle=\frac{1}{\sqrt2}\left(\ket{\frac{N}{2},\frac{N}{2}}_{x}+\ket{\frac{N}{2},-\frac{N}{2}}_{x}\right)\otimes\ket{0}_{r}$ as initial state. Acting the evolution operator in Eq.~(\ref{eq_rabi_11}) on the initial state, we obtain
\begin{equation}
\label{eq_rabi_21}
\begin{split}
|\psi(t)\rangle =\frac{1}{\sqrt2}\exp\left(iN^{2}\phi(t)/4\right)\left(\ket{\frac{N}{2},\frac{N}{2}}_{x}\otimes\ket{\frac{N}{2}\beta(t)}_{r}+\ket{\frac{N}{2},-\frac{N}{2}}_{x}\otimes\ket{-\frac{N}{2}\beta(t)}_{r}\right),
\end{split}
\end{equation}\\
where the coherent states $|\pm \frac{N}{2}\beta(t)\rangle_{r}=\hat{D}\left[\pm \frac{N}{2}\beta(t)\right]|0\rangle_{r}$ with the coherent state amplitude $\pm \frac{N}{2}\beta(t)=\pm \frac{N}{2}(\tilde{g}/\tilde{\omega}_{r})(1-e^{i\tilde{\omega}_{r}t})$. Obviously, the spin states $\ket{\frac{N}{2},-\frac{N}{2}}_{x}$ and $\ket{\frac{N}{2},\frac{N}{2}}_{x}$ undergo different dynamics, which depends on the spin states. Let us introduce the states $\ket{\pm}=\frac{1}{\sqrt{2}}\left(\ket{\frac{N}{2},\frac{N}{2}}_{x}\pm \ket{\frac{N}{2},-\frac{N}{2}}_{x}\right)$, then the evolution state can be rewritten as
\begin{eqnarray}
\label{eq_rabi_22}
 |\psi(t)\rangle &=& \frac{1}{2}\exp\left(iN^{2}\phi(t)/4\right)\left[\ket{+}\otimes\left(\ket{\frac{N}{2}\beta(t)}_{r}+\ket{-\frac{N}{2}\beta(t)}_{r}\right)+\ket{-}\otimes\left(\ket{\frac{N}{2}\beta(t)}_{r}-\ket{-\frac{N}{2}\beta(t)}_{r}\right)\right]\nonumber \\
   &=& \frac{1}{2}\exp\left(iN^{2}\phi(t)/4\right)\left[\mathcal{N}_{+}^{-1}\ket{+}\otimes\ket{{\rm Cat}_{+}(t)}_{r}+\mathcal{N}_{-}^{-1}\ket{-}\otimes\ket{{\rm Cat}_{-}(t)}_{r}\right],
\end{eqnarray}
where $\mathcal{N}_{\pm}=\left[2(1\pm\exp(-\frac{1}{2}N^{2}|\beta(t)|^{2}))\right]^{-1/2}$ and $\ket{{\rm Cat}_{\pm}}=\mathcal{N}_{\pm}\left(\ket{\frac{N}{2}\beta(t)}_{r}\pm \ket{-\frac{N}{2}\beta(t)}_{r}\right)$. The superposition coherent states $\ket{{\rm Cat}_{\pm}}$ are the so-called even and odd Schr\"{o}dinger cat states \cite{Liao2016,Haljan2005,Yin2013,Liu2005,Liao2008}. After measurement is performed on the states $\ket{+}$ and $\ket{-}$, the final state in Eq.~(\ref{eq_rabi_22}) collapses to the states $\ket{{\rm Cat}_{+}(t)}$ or $\ket{{\rm Cat}_{-}(t)}$. The probability of obtaining even and odd cat states are $\frac{1}{2}\left[1+\exp{(-\frac{1}{2}N^{2}|\beta(t)|^{2})}\right]$ and $\frac{1}{2}\left[1-\exp{(-\frac{1}{2}N^{2}|\beta(t)|^{2})}\right]$, respectively. The magnitude of the displacement $\frac{N}{2}|\beta(t)|$ changes depending on the evolution time. When $t_{0}=\pi/\tilde{\omega}_{r}$, the displacement reaches its maximum value $N\tilde{g}/\tilde{\omega}_{r}$, which indicates the maximum displacement can be enhanced by increasing number of the qubits $N$ and the relative coupling strength $\tilde{g}/\tilde{\omega}_{r}$.
 \begin{figure*}[t]
  \centering
  \includegraphics[width=0.9\textwidth]{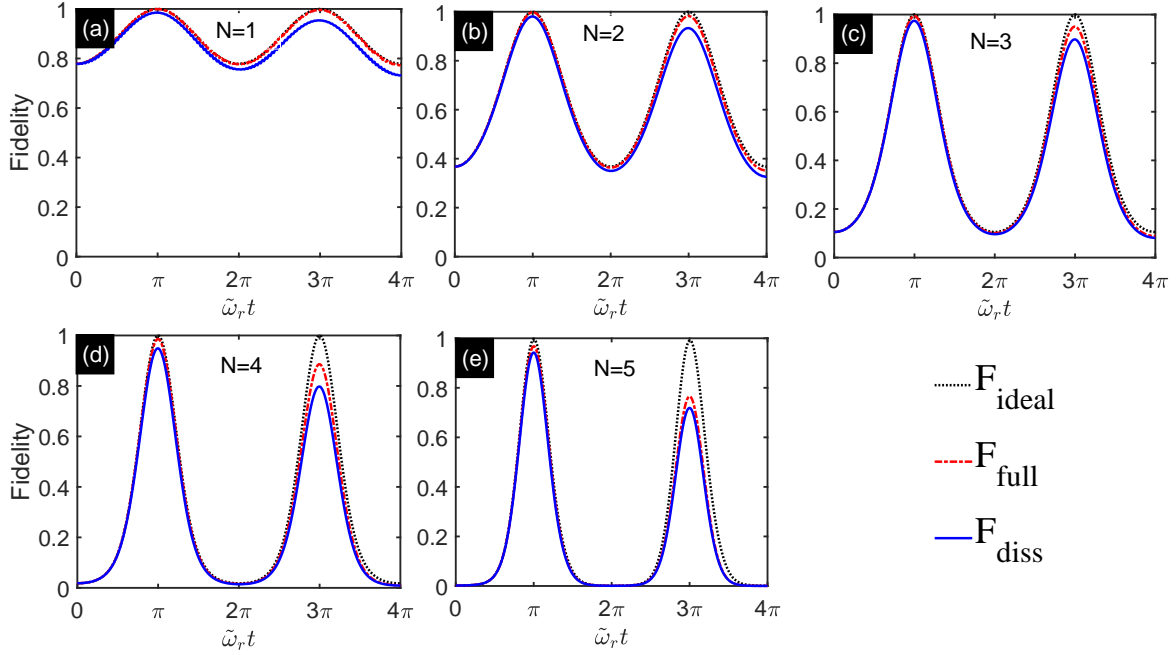}
    \caption{Numerical simulation of Schr\"{o}dinger cat states with multi-qubit system for $N=1,2,3,4,5$. The physics parameters are given in Table (\ref{tab1}). (a) $N=1$. (b) $N=2$. (c) $N=3$. (d) $N=4$. (e) $N=5$. The fidelity between target states and evolution states ${\rm F}_{\rm ideal}$, ${\rm F}_{\rm full}$, and ${\rm F}_{\rm diss}$ are plotted with black dotted line, red dash-dotted line, and blue solid line, respectively.}
  \label{fig2}
\end{figure*}

In order to study the Schr\"{o}dinger states generation when the system subjects to dissipation, we compare the evolution states under effective Hamiltonian with quantum states governed by full Hamiltonian with and without dissipation. Let $\ket{\psi(t_{0})}$ be target state. We denote time-dependent density matrix governed by the effective Hamiltonian, full Hamiltonian without dissipation and master equation with $\tilde{\rho}_{\rm ideal}(t)$, $\tilde{\rho}_{\rm full}(t)$ and $\tilde{\rho}_{\rm diss}(t)$, respectively. We compare expected state $\ket{\psi(t_{0})}$ with evolution states by using the fidelities ${\rm F}_{\rm ideal}=\bra{\psi(t_{0})}\tilde{\rho}_{\rm ideal}(t)\ket{\psi(t_{0})}$, ${\rm F}_{\rm full}=\bra{\psi(t_{0})}\tilde{\rho}_{\rm full}(t)\ket{\psi(t_{0})}$ and ${\rm F}_{\rm diss}=\bra{\psi(t_{0})}\tilde{\rho}_{\rm diss}(t)\ket{\psi(t_{0})}$. The Fig.~\ref{fig2} shows the numerical results for ${\rm F}_{\rm ideal}$ (black dotted line), ${\rm F}_{\rm full}$ (red dash-dotted line) and ${\rm F}_{\rm diss}$ (blue solid line). The results show that when evolution time $t=\pi/\tilde{\omega}_{r}$, the target state is reached. Even when the system subjects to dissipation, we also can obtain Schr\"{o}dinger cat states when the number of the qubits is not very large.

\subsection*{The generation of multi-qubit GHZ states.}
The derived effective Hamiltonian in Eq.~(\ref{eq_rabi_eff}) also can be used to generate the multi-qubit GHZ states \cite{Wang2010,MS1999}. Let $\tilde{\varepsilon}=0$ and the evolution time $t=T=2\pi/|\tilde{\omega}_{r}|$, we get $\beta(T)=0$ and $\phi(T)=2\pi \left(\tilde{g}/\tilde{\omega}_{r}\right)^{2}$. Then the evolution operator (\ref{eq_rabi_11}) reduces to
 \begin{equation}
 \label{eq_rabi_11-1}
  \mathcal{U}(T)=\exp\left(i\phi(T)\hat{J}_{x}^{2}\right).
 \end{equation}
 The multi-qubit states $\ket{gg\cdots g}$ and $\ket{ee\cdots e}$ can be recast in terms of the collective states as $\ket{\frac{N}{2},-\frac{N}{2}}_{z}$ and $\ket{\frac{N}{2},\frac{N}{2}}_{z}$, respectively. The collective states $\ket{\frac{N}{2},\pm \frac{N}{2}}_{z}$ can be expressed in terms of eigenstates of the $\{\hat{J}^{2},\hat{J}_{x}\}$ as follows
 \begin{subequations}
 \label{eq_rabi_11-2}
\begin{align}
\ket{\frac{N}{2},-\frac{N}{2}}_{z}&=\sum_{M=-N/2}^{N/2}C_{M}\ket{\frac{N}{2},M}_{x},\\
\ket{\frac{N}{2},\frac{N}{2}}_{z}&=\sum_{M=-N/2}^{N/2}C_{M}(-1)^{N/2-M}\ket{\frac{N}{2},M}_{x}.
\end{align}
\end{subequations}
Let $\ket{\psi(0)}=\ket{gg\cdots g}\equiv \ket{\frac{N}{2},-\frac{N}{2}}_{z}$ be the initial state. Acting the unitary operator (\ref{eq_rabi_11-1}) on the initial state, we obtain
 \begin{equation}
 \label{eq_rabi_11-3}
 \ket{\psi(T)}= \mathcal{U}(T)\ket{\frac{N}{2},-\frac{N}{2}}_{z}= \sum_{M=-N/2}^{N/2}C_{M}e^{i\phi(T)M^{2}}\ket{\frac{N}{2},M}_{x}.
 \end{equation}
If we set $\phi(T)=\pi/2$ (i.e., $\tilde{g}/\tilde{\omega}_{r}=1/2$), the above final state reads
 \begin{equation}
 \label{eq_rabi_11-4}
 \ket{\psi(T)}= \mathcal{U}(T)\ket{\frac{N}{2},-\frac{N}{2}}_{z}= \sum_{M=-N/2}^{N/2}C_{M}e^{i\frac{M^{2}}{2}\pi}\ket{\frac{N}{2},M}_{x}.
 \end{equation}
In the following, we proof the above final state is local equivalent to the $N-$qubit GHZ state. When $N$ is an even integer, $M$ are integers ranging from $-\frac{N}{2}$ to $\frac{N}{2}$. We also check that $e^{iM^{2}\pi/2}$ is equal to $1$ for even $M$ and $i$ for odd $M$. Then the final state in this case reads
 \begin{equation}
 \label{eq_rabi_11-5}
 \ket{\psi(T)}_{e} = \sum_{M=-N/2}^{N/2}\frac{C_{M}}{\sqrt{2}}\left(e^{i\pi/4}+(-1)^{-M}e^{-i\pi/4}\right)\ket{\frac{N}{2},M}_{x}.
 \end{equation}
 Such state can be expressed as superposition of the collective states $\ket{\frac{N}{2},\frac{N}{2}}_{z}$ and $\ket{\frac{N}{2},-\frac{N}{2}}_{z}$ as follows
\begin{eqnarray}
 \label{eq_rabi_11-6}
  \ket{\psi(T)}_{e} &=& \frac{e^{i\pi/4}}{\sqrt{2}}\left(\ket{\frac{N}{2},-\frac{N}{2}}_{z}+e^{i\left[(N-1)\pi/2)\right]}\ket{\frac{N}{2},\frac{N}{2}}_{z}\right)\nonumber \\
  &=& \frac{e^{i\pi/4}}{\sqrt{2}}\left(\ket{gg\cdots g}+e^{i\left[(N-1)\pi/2)\right]}\ket{ee\cdots e}\right)\nonumber\\
  &\equiv& \ket{{\rm GHZ}}^{(N)}_{e}.
\end{eqnarray}
 \begin{figure*}[t]
  \centering
  \includegraphics[width=0.9\textwidth]{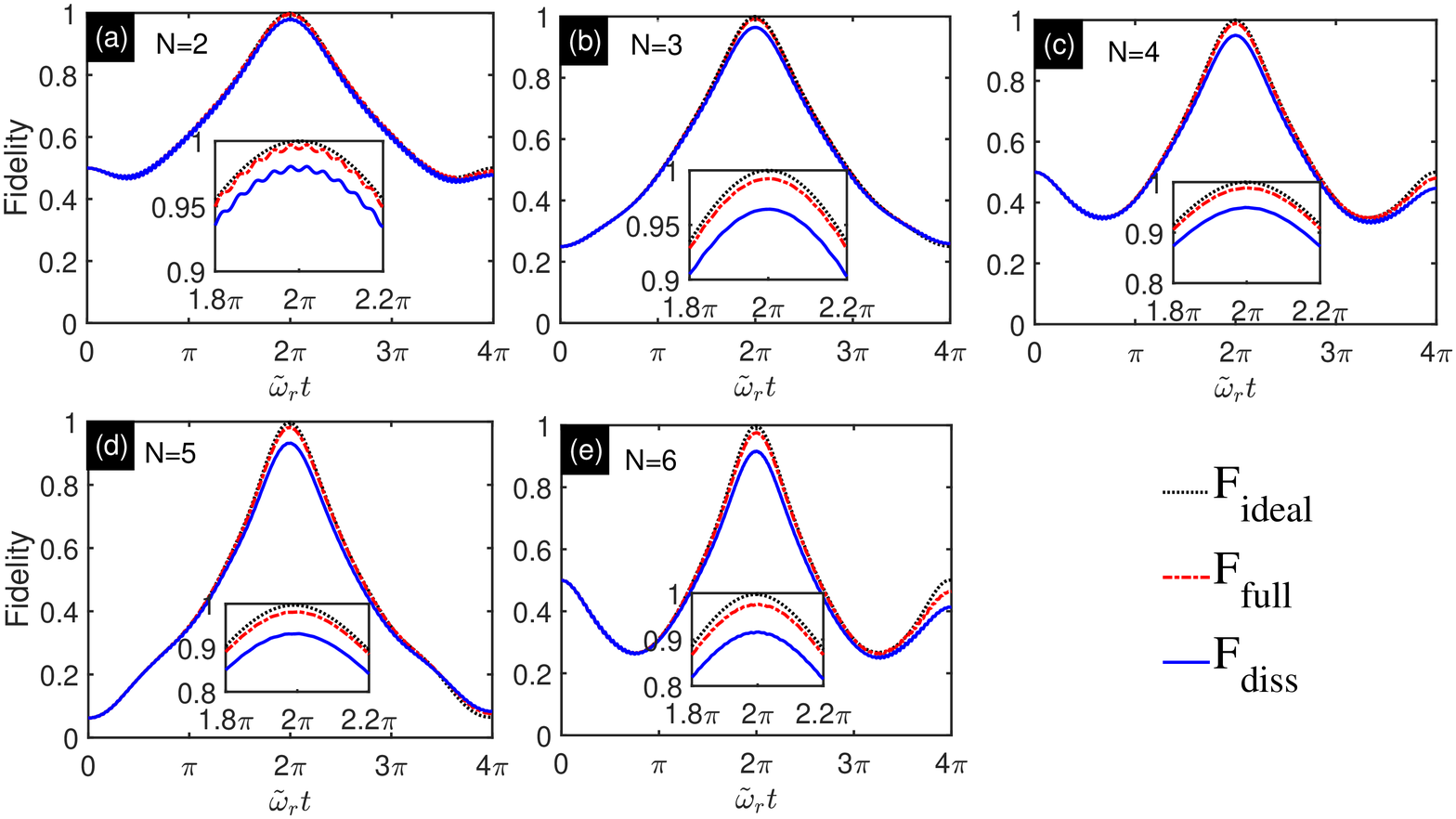}
    \caption{Numerical simulation of the multi-qubit GHZ states for $N=2,3,4,5,6$. The physics parameters are given in Table (\ref{tab1}). (a) $N=2$. (b) $N=3$. (c) $N=4$. (d) $N=5$. (e) $N=6$. The fidelity between target GHZ states and evolution states ${\rm F}_{\rm ideal}$, ${\rm F}_{\rm full}$, and ${\rm F}_{\rm diss}$ are plotted with black dotted line, red dash-dotted line, and blue solid line, respectively.}
  \label{fig3}
\end{figure*}

If $N$ is an odd integer, $M$ are half integers. We can introduce an integer $M'$ with $M'=M-\frac{1}{2}$. Then the final state (\ref{eq_rabi_11-4}) are
\begin{eqnarray}
 \label{eq_rabi_11-7}
  \ket{\psi(T)}_{o} &=& \sum_{M=-N/2}^{N/2}C_{M}e^{-i\pi/8}e^{i\frac{M}{2}\pi}e^{i\frac{M'^{2}}{2}\pi}\ket{\frac{N}{2},M}_{x}\nonumber \\
  &=& e^{-i\pi/8}e^{i\frac{\pi}{2}\hat{J}_{x}}\ket{\psi'(T)}_{o},
\end{eqnarray}
where $\ket{\psi'(T)}_{o}= \sum_{M'=-(N+1)/2}^{(N-1)/2}C_{M}e^{i\frac{M'^{2}}{2}\pi}\ket{\frac{N}{2},M}_{x}$. The state $\ket{\psi'(T)}_{o}$ is local equivalent to the state $\ket{\psi(T)}_{o}$. Considering even or odd integer $M'$, the state $\ket{\psi'(T)}_{o}$ can be rewritten as
\begin{eqnarray}
 \label{eq_rabi_11-8}
  \ket{\psi'(T)}_{o} &=& \frac{e^{i\pi/4}}{\sqrt{2}}\left(\ket{\frac{N}{2},-\frac{N}{2}}_{z}-e^{iN\pi/2}\ket{\frac{N}{2},\frac{N}{2}}_{z}\right)\nonumber \\
  &=& \frac{e^{i\pi/4}}{\sqrt{2}}\left(\ket{gg\cdots g}-e^{iN\pi/2}\ket{ee\cdots e}\right)\nonumber\\
  &\equiv& \ket{{\rm GHZ}}^{(N)}_{o}.
\end{eqnarray}
Based on the Eqs.~(\ref{eq_rabi_11-6}) and (\ref{eq_rabi_11-8}), the final state is equivalent to the GHZ state for even or odd integer $N$. The above results apply to an ideal situation, namely, dissipation-free environment. To assess the experimental feasibility of our proposal, we compare multi-qubit GHZ states $\ket{{\rm GHZ}}^{(N)}_{o/e}$ (we denote $\ket{{\rm GHZ}}$ for simplicity) with evolution states governed by the effective Hamiltonian (i.e., Eq.~(\ref{eq_rabi_eff})), the full Hamiltonian without dissipation (i.e., Eq.~(\ref{eq_rabi_1})), the full Hamiltonian with dissipation (i.e., Eq.~(\ref{eq_master_x})). We denote the evolution density matrices governed by Eq.~(\ref{eq_rabi_eff}), Eq.~(\ref{eq_rabi_1}) and Eq.~(\ref{eq_master_x}) with $\tilde{\rho}_{\rm ideal}(t)$, $\tilde{\rho}_{\rm full}(t)$ and $\tilde{\rho}_{\rm diss}(t)$, respectively. The fidelity between multi-qubit GHZ states $\ket{{\rm GHZ}}$ and evolution states are denoted by ${\rm F}_{\rm ideal}= \bra{\rm GHZ}\tilde{\rho}_{\rm ideal}(t)\ket{{\rm GHZ}}$, ${\rm F}_{\rm full}= \bra{\rm GHZ}\tilde{\rho}_{\rm full}(t)\ket{{\rm GHZ}}$ and ${\rm F}_{\rm diss}= \bra{\rm GHZ}\tilde{\rho}_{\rm diss}(t)\ket{{\rm GHZ}}$. The Fig.~\ref{fig3} shows the numerical results for ${\rm F}_{\rm ideal}$ (black dotted line), ${\rm F}_{\rm full}$ (red dash-dotted line) and ${\rm F}_{\rm diss}$ (blue solid line). The Fig.~\ref{fig3} (a)-\ref{fig3}(e) are the numerical results for $N=2,3,4,5,6$, respectively. The fidelity for ${\rm F}_{\rm full}$ and ${\rm F}_{\rm diss}$ at time $t=2\pi/\tilde{\omega}_{r}$ are shown in Table (\ref{tab2}). The results show that we can obtain high fidelity multi-qubit GHZ state even the system subjecting to dissipation.
\begin{table}
\centering
\caption{The fidelity of the GHZ states at time $t=2\pi/\tilde{\omega}_{r}$ is list in the following table.}
\label{tab2}
\begin{tabular}{cccccc}
\hline
\hline
 & $N=2$ & $N=3$ & $N=4$ & $N=5$ & $N=6$\\
\hline
${\rm F}_{\rm full}$ & 0.9971 & 0.9924 & 0.9886 & 0.9816 & 0.9761\\
\hline
${\rm F}_{\rm diss}$ & 0.9806 & 0.9644 & 0.9497 & 0.9321 & 0.9162\\
\hline
\hline
\end{tabular}
\end{table}

\section*{Discussion}

In summery, we have proposed a scheme to simulate the multi-qubit quantum Rabi model in circuit QED system. The effective Hamiltonian for multi-qubit quantum Rabi model can be derived. Based on unitary dynamics, the fidelity of effective Hamiltonian is discussed in detail. The results show that the system can reach stronger coupling regimes by adjusting the external driving amplitudes and frequencies. With this tunable effective Hamiltonian, the qubit-dependent displacement interaction Hamiltonian can be obtained by tuning the driving parameters. Based on such Hamiltonian, we also discuss the applications to constructing nontrivial quantum gate, the Schr\"{o}dinger cat states and multi-qubit GHZ states. With the effective Hamiltonian, we can generate the quantum gate with the maximum two-qubit entangling power. The local equivalence between the achieved quantum gate and the CNOT gate has been discussed in detail. The numerical calculation shows that the process fidelity of the quantum gate reaches $96.32\%$ under the chosen parameters. The Schr\"{o}dinger cat states can be generated with the effective Hamiltonian, and the magnitude of the displacement can be enhanced by increasing the number of the qubits and relative coupling strength. In the case of multiple quantum qubits, we generate high fidelity multi-qubit GHZ states for even and odd $N$. We show that the high fidelity Schr\"{o}dinger cat state and multi-qubit GHZ state can be obtained even the system subjecting to dissipation.

The presented proposal may open a way to study the stronger coupling regimes whose coupling strength is far away from ultrastrong coupling regimes. We should note that the effective Hamiltonian is not validity when the number of the qubits and the relative coupling strength are very large. Even so, our scheme may also provide potential applications to the quantum computation and quantum state engineering.

\section*{Methods}
In this part, we will show how to obtain the effective Hamiltonian in Eq.~(\ref{eq_rabi_eff}). We choose the rotating framework related to the time-dependent unitary transformation $U_{1}(t)=e^ {-i\omega_{x}\hat{J}_{z}t}e^{-i\omega_{x} \hat{a}^{\dag}\hat{a}t}$. The transformed Hamiltonian reads
\begin{eqnarray}
\label{eq_rabi_3}
  \hat{H}'(t) &=& U_{1}^{\dag}(t)\hat{H}U_{1}(t)-iU_{1}^{\dag}(t)\frac{\partial U_{1}(t)}{\partial t}\nonumber\\
              &=& (\omega_{r}-\omega_{x})\hat{a}^{\dag}\hat{a} + [\varepsilon-\omega_{x}+\Omega_{z} \cos(\omega_{z} t)]\hat{J}_{z}\nonumber\\
              && +
   \frac{\Omega_{x}}{4}\left[\hat{J}_{+}(1+e^{2i\omega_{x} t})+{\rm H.c.}\right] + \frac{g}{2}(\hat{a}^{\dag}\hat{J}_{-}+\hat{a}\hat{J}_{-}e^{-2i\omega_{x}t}+{\rm H.c.}),
\end{eqnarray}
where $\hat{J}_{\pm}=\hat{J}_{x}\pm i\hat{J}_{y}$ and ${\rm H.c.}$ denotes the Hermitian conjugate. Considering the condition $\omega_{x}\gg \Omega_{x}$ and $\omega_{x}\gg g$, the fast oscillating terms can be ignored by performing RWA. Then the simplified Hamiltonian reads
 \begin{equation}
\label{eq_rabi_3-2}
  \hat{H}'(t) \approx (\omega_{r}-\omega_{x})\hat{a}^{\dag}\hat{a}+[\varepsilon-\omega_{x}+\Omega_{z} \cos(\omega_{z} t)]\hat{J}_{z} +
   \frac{\Omega_{x}}{2}\hat{J}_{x}+\frac{g}{2}(\hat{a}^{\dag}\hat{J}_{-}+
  \hat{a}\hat{J}_{+}),
\end{equation}
Moving to the rotation framework with respect to $U_{2}(t)=e^{-i\hat{H}'_{0}t}$ with $\hat{H}'_{0}=\frac{\Omega_{x}}{2}\hat{J}_{x}$, we obtain the following transformed Hamiltonian
\begin{eqnarray}
\label{eq_rabi_4}
  \hat{H}''(t) &=& U_{2}^{\dag}(t)\hat{H}'U_{2}(t)-iU_{2}^{\dag}(t)\frac{\partial U_{2}(t)}{\partial t}\nonumber\\
   &=&(\omega_{r}-\omega_{x})\hat{a}^{\dag}\hat{a}+(\varepsilon-\omega_{x}+\Omega_{z}\cos(\omega_{z} t))\left[\cos\left(\frac{\Omega_{x}}{2}t\right)\hat{J}_{z}+\sin\left(\frac{\Omega_{x}}{2}t\right)\hat{J}_{y}\right]\nonumber\\
   && + g\left\{\left[i\sin\left(\frac{\Omega_{x}}{2}t\right)\hat{J}_{z}+\cos^{2}\left(\frac{\Omega_{x}t}{4}\right)\hat{J}_{-}+\sin^{2}\left(\frac{\Omega_{x}t}{4}\right)\hat{J}_{+}\right]\hat{a}^{\dag}+{\rm H.c.}\right\}.
\end{eqnarray}
Let $\Omega_{x}=2\omega_{z}$. The Hamiltonian in Eq.~(\ref{eq_rabi_4}) can be rewritten as
\begin{eqnarray}
\label{eq_rabi_4-1}
  \hat{H}'''(t) &=& (\omega_{r}-\omega_{x})\hat{a}^{\dag}\hat{a} +(\varepsilon-\omega_{x})[\cos(\omega_{z}t)\hat{J}_{z}+\sin(\omega_{z}t)\hat{J}_{y}]+\frac{\Omega_{z}}{2}[(1+\cos(2\omega_{z}t))\hat{J}_{z}+\sin(2\omega_{z}t)\hat{J}_{y}]\nonumber\\
   &&+ \frac{g}{2}\left[i\sin(\omega_{z}t)\hat{J}_{z}\hat{a}^{\dag}+\frac{1}{2}(1+\cos\left(\omega_{z}t\right))\hat{J}_{-}\hat{a}^{\dag}+\frac{1}{2}(1-\cos\left(\omega_{z}t\right))\hat{J}_{+}\hat{a}^{\dag}+{\rm H.c.}\right].
\end{eqnarray}
When the parameters satisfy the conditions: $\omega_{z} \gg \Omega_{z}$, $\omega_{z}\gg\varepsilon-\omega_{x}$, $\omega_{z}\gg g$, we can neglect the fast oscillating terms and recast above Hamiltonian as
\begin{equation}
\label{eq_rabi_5x}
  \hat{H}_{\rm eff}=\tilde{\omega}_{r}\hat{a}^{\dag}\hat{a}+\tilde{\varepsilon}\hat{J}_{z}+
\tilde{g}(\hat{a}+\hat{a}^{\dag})\hat{J}_{x},
\end{equation}
where $\tilde{\omega}_{r}=\omega_{r}-\omega_{x}$, $\tilde{\varepsilon}=\Omega_{z}/2$, and $\tilde{g}=g/2$. Thus we obtain the effective Hamiltonian shown in Eq.~(\ref{eq_rabi_eff}), which is effective under the following conditions: $\Omega_{x}=2\omega_{z}$, $\omega_{x}\gg \Omega_{x} \gg g$, $\omega_{z}\gg\Omega_{z}$, $\omega_{z}\gg\varepsilon-\omega_{x}$. Such Hamiltonian is multi-qubit extension of the QRM with tunable parameters (i.e., the tunable Dicke model). The simulated coupling ratio is  $\tilde{g}/\tilde{\omega}_{r}=g/[2(\omega_{r}-\omega_{x})]$, which is also turnable by adjusting the frequency of the transverse driving.


\begin{thebibliography}{10}
\expandafter\ifx\csname url\endcsname\relax
  \def\url#1{\texttt{#1}}\fi
\expandafter\ifx\csname urlprefix\endcsname\relax\def\urlprefix{URL }\fi
\providecommand{\bibinfo}[2]{#2}
\providecommand{\eprint}[2][]{\url{#2}}

\bibitem{Rabi1936}
\bibinfo{author}{Rabi, I. I.}
\newblock \bibinfo{title}{On the Process of Space Quantization}.
\newblock \emph{\bibinfo{journal}{Physical Review}}
  \textbf{\bibinfo{volume}{49}}, \bibinfo{pages}{324}
  (\bibinfo{year}{1936}).

\bibitem{Rabi1937}
\bibinfo{author}{Rabi, I. I.}
\newblock \bibinfo{title}{Space Quantization in a Gyrating Magnetic Field}.
\newblock \emph{\bibinfo{journal}{Physical Review}}
  \textbf{\bibinfo{volume}{51}}, \bibinfo{pages}{652}
  (\bibinfo{year}{1937}).

\bibitem{braak2011}
\bibinfo{author}{Braak, D.}
\newblock \bibinfo{title}{Integrability of the Rabi Model}.
\newblock \emph{\bibinfo{journal}{Phys. Rev. Lett.}}
  \textbf{\bibinfo{volume}{107}}, \bibinfo{pages}{100401}
  (\bibinfo{year}{2011}).

\bibitem{scully}
\bibinfo{author}{Scully, M.-O.} \& \bibinfo{author}{Zubairy, M.-S.}
\newblock \emph{\bibinfo{title}{Quantum Optics}}
  (\bibinfo{publisher}{Cambridge university press}, \bibinfo{year}{1997}).

\bibitem{irish2007}
\bibinfo{author}{Irish, E.-K.}
\newblock \bibinfo{title}{Generalized Rotating-Wave Approximation for Arbitrarily Large Coupling}.
\newblock \emph{\bibinfo{journal}{Phys. Rev. Lett.}}
  \textbf{\bibinfo{volume}{99}}, \bibinfo{pages}{173601}
  (\bibinfo{year}{2007}).

\bibitem{thanopulos2004}
\bibinfo{author}{Thanopulos, I.}, \bibinfo{author}{Paspalakis, E.} \& \bibinfo{author}{Kis, Z.}
\newblock \bibinfo{title}{Laser-driven coherent manipulation of molecular chirality}.
\newblock \emph{\bibinfo{journal}{Chem. Phys. Lett.}}
  \textbf{\bibinfo{volume}{390}}, \bibinfo{pages}{228}
  (\bibinfo{year}{2004}).

\bibitem{Jaynes1963}
\bibinfo{author}{Jaynes, E.-T.} \& \bibinfo{author}{Cummings, F.-W.}
\newblock \bibinfo{title}{Comparison of quantum and semiclassical radiation theories with application to the beam maser}.
\newblock \emph{\bibinfo{journal}{Proc. IEEE}}
  \textbf{\bibinfo{volume}{51}}, \bibinfo{pages}{89}
  (\bibinfo{year}{1963}).

\bibitem{Cummings1965}
\bibinfo{author}{Cummings, F.-W.}
\newblock \bibinfo{title}{Stimulated emission of radiation in a single
mode}.
\newblock \emph{\bibinfo{journal}{Physical Review}}
  \textbf{\bibinfo{volume}{140}}, \bibinfo{pages}{A1051-A1056}
  (\bibinfo{year}{1965}).

\bibitem{Eberly1980}
\bibinfo{author}{Eberly, J.-H.}, \bibinfo{author}{Narozhny, N.-B.} \& \bibinfo{author}{Sanchez-Mondragon, J.-J.}
\newblock \bibinfo{title}{Periodic spontaneous collapse and revival in a simple quantum model}.
\newblock \emph{\bibinfo{journal}{Phys. Rev. Lett.}}
  \textbf{\bibinfo{volume}{44}}, \bibinfo{pages}{1323}
  (\bibinfo{year}{1980}).

\bibitem{Thompson1992}
\bibinfo{author}{Thompson, R.}, \bibinfo{author}{Rempe, G.} \& \bibinfo{author}{Kimble, H.-J.}
\newblock \bibinfo{title}{Observation of normal-mode splitting for an atom in an optical cavity}.
\newblock \emph{\bibinfo{journal}{Phys. Rev. Lett.}}
  \textbf{\bibinfo{volume}{68}}, \bibinfo{pages}{1132}
  (\bibinfo{year}{1992}).

\bibitem{Boca2004}
\bibinfo{author}{Boca, A.}, \bibinfo{author}{Miller, R.}, \bibinfo{author}{Birnbaum, K.-M.}, \bibinfo{author}{Boozer, A.-D.}, \bibinfo{author}{McKeever, J.} \& \bibinfo{author}{Kimble, H.-J.}
\newblock \bibinfo{title}{Observation of the
Vacuum Rabi Spectrum for One Trapped Atom}.
\newblock \emph{\bibinfo{journal}{Phys. Rev. Lett.}}
  \textbf{\bibinfo{volume}{93}}, \bibinfo{pages}{233603}
  (\bibinfo{year}{2004}).

\bibitem{Michel2007}
\bibinfo{author}{Devoret, M.-H.}, \bibinfo{author}{Girvin, S.} \& \bibinfo{author}{Schoelkopf, R.}
\newblock \bibinfo{title}{Circuit-QED: How strong can the coupling between a Josephson junction atom and a transmission line resonator be?}.
\newblock \emph{\bibinfo{journal}{Annalen der Physik}}
  \textbf{\bibinfo{volume}{16}}, \bibinfo{pages}{767}
  (\bibinfo{year}{2007}).

\bibitem{Bourassa2009}
\bibinfo{author}{Bourassa, J.}, \bibinfo{author}{Gambetta, J.-M.}, \bibinfo{author}{Abdumalikov, A.-A.}, \bibinfo{author}{Astafiev, O.}, \bibinfo{author}{Nakamura, Y.} \& \bibinfo{author}{Blais, A.}
\newblock \bibinfo{title}{Ultrastrong coupling regime of cavity QED with phase-biased flux qubits}.
\newblock \emph{\bibinfo{journal}{Phys. Rev. A}}
  \textbf{\bibinfo{volume}{80}}, \bibinfo{pages}{032109}
  (\bibinfo{year}{2009}).

\bibitem{Casanova2010}
\bibinfo{author}{Casanova, J.}, \bibinfo{author}{Romero, G.}, \bibinfo{author}{Lizuain, I.}, \bibinfo{author}{Garc\'{\i}a-Ripoll, J.-J.} \& \bibinfo{author}{Solano, E.}
\newblock \bibinfo{title}{Deep Strong Coupling Regime of the Jaynes-Cummings Model}.
\newblock \emph{\bibinfo{journal}{Phys. Rev. Lett.}}
  \textbf{\bibinfo{volume}{105}}, \bibinfo{pages}{263603}
  (\bibinfo{year}{2010}).

\bibitem{Wallraff2004}
\bibinfo{author}{Wallraff, A.}, \bibinfo{author}{Schuster, D.-I.}, \bibinfo{author}{Blais, A.}, \bibinfo{author}{Frunzio, L.}, \bibinfo{author}{Huang, R.-S.}, \bibinfo{author}{Majer, J.}, \bibinfo{author}{Kumar, S.}, \bibinfo{author}{Girvin, S.-M.} \& \bibinfo{author}{Schoelkopf, R.-J.}
\newblock \bibinfo{title}{Strong coupling of a single photon to a superconducting qubit using circuit quantum electrodynamics}.
\newblock \emph{\bibinfo{journal}{Nature}}
  \textbf{\bibinfo{volume}{431}}, \bibinfo{pages}{162}
  (\bibinfo{year}{2004}).

\bibitem{Srinivasan2011}
\bibinfo{author}{Srinivasan, S.-J.}, \bibinfo{author}{Hoffman, A.-J.}, \bibinfo{author}{Gambetta, J.-M.} \& \bibinfo{author}{Houck, A. A.}
\newblock \bibinfo{title}{Tunable Coupling in Circuit Quantum Electrodynamics Using a Superconducting Charge Qubit with a $V$-Shaped Energy Level Diagram}.
\newblock \emph{\bibinfo{journal}{Phys. Rev. Lett.}}
  \textbf{\bibinfo{volume}{106}}, \bibinfo{pages}{083601}
  (\bibinfo{year}{2011}).

\bibitem{Niemczyk2010}
\bibinfo{author}{Niemczyk, T.}, \bibinfo{author}{Deppe, F.}, \bibinfo{author}{Huebl, H.}, \bibinfo{author}{Menzel, E.-P.}, \bibinfo{author}{Hocke, F.}, \bibinfo{author}{Schwarz, M.-J.}, \bibinfo{author}{Garcia-Ripoll, J. J.}, \bibinfo{author}{Zueco, D.}, \bibinfo{author}{H\"{u}mmer, T.}, \bibinfo{author}{Solano, E.}, \bibinfo{author}{Marx, A.} \& \bibinfo{author}{Gross, R.}
\newblock \bibinfo{title}{Circuit quantum electrodynamics in the ultrastrong-coupling regime}.
\newblock \emph{\bibinfo{journal}{Nature Physics}}
  \textbf{\bibinfo{volume}{6}}, \bibinfo{pages}{772}
  (\bibinfo{year}{2010}).

\bibitem{garz2016}
\bibinfo{author}{Garziano, L.}, \bibinfo{author}{Macr\'i, V.}, \bibinfo{author}{Stassi, R.}, \bibinfo{author}{Stefano, O.-D.}, \bibinfo{author}{Nori, F.} \& \bibinfo{author}{Savasta, S.}
\newblock \bibinfo{title}{One photon can simultaneously excite two or more atoms}.
\newblock \emph{\bibinfo{journal}{Phys. Rev. Lett.}}
  \textbf{\bibinfo{volume}{117}}, \bibinfo{pages}{043601}
  (\bibinfo{year}{2016}).

\bibitem{wangx2017}
\bibinfo{author}{Wang, X.}, \bibinfo{author}{Miranowicz, A.}, \bibinfo{author}{Li, H.-R.} \& \bibinfo{author}{Nori, F.}
\newblock \bibinfo{title}{Observing pure effects of counter-rotating terms without ultrastrong coupling: A single photon can simultaneously excite two qubits}.
\newblock \emph{\bibinfo{journal}{Phys. Rev. A}}
 \textbf{\bibinfo{volume}{96}}, \bibinfo{pages}{063820}
 (\bibinfo{year}{2017}).

\bibitem{Ridolfo2012}
\bibinfo{author}{Ridolfo, A.}, \bibinfo{author}{Leib, M.}, \bibinfo{author}{Savasta, S.} \& \bibinfo{author}{Hartmann, M. J.}
\newblock \bibinfo{title}{Photon Blockade in the Ultrastrong Coupling Regime}.
\newblock \emph{\bibinfo{journal}{Phys. Rev. Lett.}}
 \textbf{\bibinfo{volume}{109}}, \bibinfo{pages}{193602}
 (\bibinfo{year}{2012}).

\bibitem{Ridolfo2013}
\bibinfo{author}{Ridolfo, A.}, \bibinfo{author}{Savasta, S.} \& \bibinfo{author}{Hartmann, M. J.}
\newblock \bibinfo{title}{Nonclassical Radiation from Thermal Cavities in the Ultrastrong Coupling Regime}.
\newblock \emph{\bibinfo{journal}{Phys. Rev. Lett.}}
 \textbf{\bibinfo{volume}{110}}, \bibinfo{pages}{163601}
 (\bibinfo{year}{2013}).

\bibitem{Law2013}
\bibinfo{author}{Law, C.-K.}
\newblock \bibinfo{title}{Vacuum Rabi oscillation induced by virtual photons in the ultrastrong-coupling regime}.
\newblock \emph{\bibinfo{journal}{Phys. Rev. A}}
 \textbf{\bibinfo{volume}{87}}, \bibinfo{pages}{045804}
 (\bibinfo{year}{2013}).

\bibitem{caox2010}
\bibinfo{author}{Cao, X.-F.}, \bibinfo{author}{You, J.-Q.}, \bibinfo{author}{Zheng, H.}, \bibinfo{author}{Kofman, A.-G.} \& \bibinfo{author}{Nori, F.}
\newblock \bibinfo{title}{Dynamics and quantum Zeno effect for a qubit in either a low- or high-frequency bath beyond the rotating-wave approximation}.
\newblock \emph{\bibinfo{journal}{Phys. Rev. A}}
 \textbf{\bibinfo{volume}{82}}, \bibinfo{pages}{022119}
 (\bibinfo{year}{2010}).

\bibitem{aiq2010}
\bibinfo{author}{Ai, Q.}, \bibinfo{author}{Li, Y.}, \bibinfo{author}{Zheng, H.} \& \bibinfo{author}{Sun, C.-P.}
\newblock \bibinfo{title}{Quantum anti-Zeno effect without rotating wave approximation}.
\newblock \emph{\bibinfo{journal}{Phys. Rev. A}}
 \textbf{\bibinfo{volume}{81}}, \bibinfo{pages}{042116}
 (\bibinfo{year}{2010}).

\bibitem{lipb2012}
\bibinfo{author}{Li, P.-B.}, \bibinfo{author}{Gao, S.-Y.} \& \bibinfo{author}{Li, F. L.}
\newblock \bibinfo{title}{Engineering two-mode entangled states between two superconducting resonators by dissipation}.
\newblock \emph{\bibinfo{journal}{Phys. Rev. A}}
 \textbf{\bibinfo{volume}{86}}, \bibinfo{pages}{012318}
 (\bibinfo{year}{2012}).

\bibitem{wangx2014}
\bibinfo{author}{Wang, X.}, \bibinfo{author}{Li, H.-R.}, \bibinfo{author}{Li, P.-B.}, \bibinfo{author}{Jiang, C.-W.}, \bibinfo{author}{Gao, H.} \& \bibinfo{author}{Li, F. L.}
\newblock \bibinfo{title}{Preparing ground states and squeezed states of nanomechanical cantilevers by fast dissipation}.
\newblock \emph{\bibinfo{journal}{Phys. Rev. A}}
  \textbf{\bibinfo{volume}{90}}, \bibinfo{pages}{013838}
  (\bibinfo{year}{2014}).

\bibitem{reiter2013}
\bibinfo{author}{Reiter, F.}, \bibinfo{author}{Tornberg, L.}, \bibinfo{author}{Johansson, G.} \& \bibinfo{author}{S{\o}rensen, A. S.}
\newblock \bibinfo{title}{Steady-state entanglement of two superconducting qubits engineered by dissipation}.
\newblock \emph{\bibinfo{journal}{Phys. Rev. A}}
  \textbf{\bibinfo{volume}{88}}, \bibinfo{pages}{032317}
  (\bibinfo{year}{2013}).

\bibitem{hes2014}
\bibinfo{author}{He, S.}, \bibinfo{author}{Zhao, Y.} \& \bibinfo{author}{Chen, Q.-H.}
\newblock \bibinfo{title}{Absence of collapse in quantum Rabi oscillations}.
\newblock \emph{\bibinfo{journal}{Phys. Rev. A}}
  \textbf{\bibinfo{volume}{90}}, \bibinfo{pages}{053848}
  (\bibinfo{year}{2014}).


\bibitem{Huang2017}
\bibinfo{author}{Huang, J.-F.}, \bibinfo{author}{Liao, J.-Q.}, \bibinfo{author}{Tian, L.} \& \bibinfo{author}{Kuang, L.-M.}
\newblock \bibinfo{title}{Manipulating counter-rotating interactions in the quantum Rabi model via modulation of the transition frequency of the two-level system}.
\newblock \emph{\bibinfo{journal}{Phys. Rev. A}}
  \textbf{\bibinfo{volume}{96}}, \bibinfo{pages}{043849}
  (\bibinfo{year}{2017}).



\bibitem{rossatto2016}
\bibinfo{author}{Rossatto, D.-Z.}, \bibinfo{author}{Felicetti, S.}, \bibinfo{author}{Eneriz, H.}, \bibinfo{author}{Rico, E.}, \bibinfo{author}{Sanz, M.} \& \bibinfo{author}{Solano, E.}
\newblock \bibinfo{title}{Entangling polaritons via dynamical Casimir effect in circuit quantum electrodynamics}.
\newblock \emph{\bibinfo{journal}{Phys. Rev. B}}
  \textbf{\bibinfo{volume}{93}}, \bibinfo{pages}{094514}
  (\bibinfo{year}{2016}).

\bibitem{felicettiprl2014}
\bibinfo{author}{Felicetti, S.}, \bibinfo{author}{Sanz, M.}, \bibinfo{author}{Lamata, L.}, \bibinfo{author}{Romero, G.}, \bibinfo{author}{Johansson, G.}, \bibinfo{author}{Delsing, P.} \& \bibinfo{author}{Solano, E.}
\newblock \bibinfo{title}{Dynamical Casimir Effect Entangles Artificial Atoms}.
\newblock \emph{\bibinfo{journal}{Phys. Rev. Lett.}}
  \textbf{\bibinfo{volume}{113}}, \bibinfo{pages}{093602}
  (\bibinfo{year}{2014}).

\bibitem{kyawprb2015}
\bibinfo{author}{Kyaw, T. H.}, \bibinfo{author}{Herrera-Mart\'i, D. A.}, \bibinfo{author}{Solano, E.}, \bibinfo{author}{Romero, G.} \& \bibinfo{author}{Kwek, L.-C.}
\newblock \bibinfo{title}{Creation of quantum error correcting codes in the ultrastrong coupling regime}.
\newblock \emph{\bibinfo{journal}{Phys. Rev. B}}
  \textbf{\bibinfo{volume}{91}}, \bibinfo{pages}{064503}
  (\bibinfo{year}{2015}).

\bibitem{romero2012}
\bibinfo{author}{Romero, G.}, \bibinfo{author}{Ballester, D.}, \bibinfo{author}{Wang, Y.-M.}, \bibinfo{author}{Scarani, V.} \& \bibinfo{author}{Solano, E.}
\newblock \bibinfo{title}{Ultrafast Quantum Gates in Circuit QED}.
\newblock \emph{\bibinfo{journal}{Phys. Rev. Lett.}}
  \textbf{\bibinfo{volume}{108}}, \bibinfo{pages}{120501}
  (\bibinfo{year}{2012}).

\bibitem{wangym2017}
\bibinfo{author}{Wang, Y.-M.}, \bibinfo{author}{Guo, C.}, \bibinfo{author}{Zhang, G.-Q.}, \bibinfo{author}{Wang, G. C.} \& \bibinfo{author}{Wu, C. F.}
\newblock \bibinfo{title}{Ultrafast quantum computation in ultrastrongly coupled circuit QED systems}.
\newblock \emph{\bibinfo{journal}{Sci. Rep.}}
  \textbf{\bibinfo{volume}{7}}, \bibinfo{pages}{44251}
  (\bibinfo{year}{2017}).

\bibitem{cui2018}
\bibinfo{author}{Cui, X.}, \bibinfo{author}{Wang, Z.} \& \bibinfo{author}{Li, Y.}
\newblock \bibinfo{title}{Detection of emitter-resonator coupling strength in the quantum Rabi model via an auxiliary resonator}.
\newblock \emph{\bibinfo{journal}{Phys. Rev. A}}
  \textbf{\bibinfo{volume}{98}}, \bibinfo{pages}{043812}
  (\bibinfo{year}{2018}).


\bibitem{Deng2015}
\bibinfo{author}{Deng C.}, \bibinfo{author}{Orgiazzi, J.}, \bibinfo{author}{Shen, F.}, \bibinfo{author}{Ashhab, S.} \& \bibinfo{author}{Lupascu, A.}
\newblock \bibinfo{title}{Observation of Floquet States in a Strongly Driven Artificial Atom}.
\newblock \emph{\bibinfo{journal}{Phys. Rev. Lett.}}
  \textbf{\bibinfo{volume}{115}}, \bibinfo{pages}{133601}
  (\bibinfo{year}{2015}).

\bibitem{Ballester2012}
\bibinfo{author}{Ballester, D.}, \bibinfo{author}{Romero, G.}, \bibinfo{author}{Garc\'{\i}a-Ripoll, J.-J.}, \bibinfo{author}{Deppe, F.} \& \bibinfo{author}{Solano, E.}
\newblock \bibinfo{title}{Quantum Simulation of the Ultrastrong Coupling Dynamics in Circuit QED}.
\newblock \emph{\bibinfo{journal}{Phys. Rev. X}}
  \textbf{\bibinfo{volume}{2}}, \bibinfo{pages}{021007}
  (\bibinfo{year}{2012}).

\bibitem{Li2013}
\bibinfo{author}{Li, J.}, \bibinfo{author}{Silveri, M.-P.}, \bibinfo{author}{Kumar, K.-S.}, \bibinfo{author}{Pirkkalainen, J.-M.}, \bibinfo{author}{Veps\"{a}l\"{a}inen, A.}, \bibinfo{author}{Chien, W.-C.}, \bibinfo{author}{Tuorila, J.}, \bibinfo{author}{Sillanp\"{a}\"{a}, M.-A.}, \bibinfo{author}{Hakonen, P.-J.}, \bibinfo{author}{Thuneberg, E.-V.} \& \bibinfo{author}{Paraoanu, G.-S.}
\newblock \bibinfo{title}{Motional averaging in a superconducting qubit}.
\newblock \emph{\bibinfo{journal}{Nature Communications}}
  \textbf{\bibinfo{volume}{4}}, \bibinfo{pages}{1420}
  (\bibinfo{year}{2013}).

\bibitem{Braumuller2017}
\bibinfo{author}{Braum\"{u}ller, J.}, \bibinfo{author}{Marthaler, M.}, \bibinfo{author}{Schneider, A.}, \bibinfo{author}{Stehli, A.}, \bibinfo{author}{Rotzinger, H.}, \bibinfo{author}{Weides, M.} \& \bibinfo{author}{Ustinov, A.-V.}
\newblock \bibinfo{title}{Analog quantum simulation of the Rabi model in the ultra-strong coupling regime}.
\newblock \emph{\bibinfo{journal}{Nature Communications}}
  \textbf{\bibinfo{volume}{8}}, \bibinfo{pages}{779}
  (\bibinfo{year}{2017}).

\bibitem{wangym2018}
\bibinfo{author}{Wang, Y.-M.}, \bibinfo{author}{You, W.-L.}, \bibinfo{author}{Liu, M.-X.}, \bibinfo{author}{Dong, Y.-L.}, \bibinfo{author}{Luo, H.-G.}, \bibinfo{author}{Romero, G.} \& \bibinfo{author}{You, J.-Q.}
\newblock \bibinfo{title}{Quantum criticality and state engineering in the simulated anisotropic quantum Rabi model}.
\newblock \emph{\bibinfo{journal}{New J. Phys.}}
  \textbf{\bibinfo{volume}{20}}, \bibinfo{pages}{053061}
  (\bibinfo{year}{2018}).


\bibitem{Crespi2012}
\bibinfo{author}{Crespi, A.}, \bibinfo{author}{Longhi, S.} \& \bibinfo{author}{Osellame, R.}
\newblock \bibinfo{title}{Photonic Realization of the Quantum Rabi Model}.
\newblock \emph{\bibinfo{journal}{Phys. Rev. Lett.}}
  \textbf{\bibinfo{volume}{108}}, \bibinfo{pages}{163601}
  (\bibinfo{year}{2012}).

\bibitem{Pedernales2015}
\bibinfo{author}{Pedernales, J.-S.}, \bibinfo{author}{Lizuain, I.}, \bibinfo{author}{Felicetti, S.}, \bibinfo{author}{Romero, G.}, \bibinfo{author}{Lamata, L.} \& \bibinfo{author}{Solano, E.}
\newblock \bibinfo{title}{Quantum Rabi Model with Trapped Ions}.
\newblock \emph{\bibinfo{journal}{Sci. Rep.}}
  \textbf{\bibinfo{volume}{5}}, \bibinfo{pages}{15472}
  (\bibinfo{year}{2015}).

\bibitem{Aedo2018}
\bibinfo{author}{Aedo, I.} \& \bibinfo{author}{Lamata, L.}
\newblock \bibinfo{title}{Analog quantum simulation of generalized Dicke models in trapped ions}.
\newblock \emph{\bibinfo{journal}{Phys. Rev. A}}
  \textbf{\bibinfo{volume}{97}}, \bibinfo{pages}{042317}
  (\bibinfo{year}{2018}).

\bibitem{Felicetti2017}
\bibinfo{author}{Felicetti, S.}, \bibinfo{author}{Romero, G.}, \bibinfo{author}{Solano, E.} \& \bibinfo{author}{Sab\'{\i}n, C.}
\newblock \bibinfo{title}{Quantum Rabi model in a superfluid Bose-Einstein condensate}.
\newblock \emph{\bibinfo{journal}{Phys. Rev. A}}
  \textbf{\bibinfo{volume}{96}}, \bibinfo{pages}{033839}
  (\bibinfo{year}{2017}).

\bibitem{Felicetti2017_2}
\bibinfo{author}{Felicetti, S.}, \bibinfo{author}{Rico, E.}, \bibinfo{author}{Sab\'{\i}n, C.}, \bibinfo{author}{Ockenfels, T.},
 \bibinfo{author}{Ockenfels, T.}, \bibinfo{author}{Koch, J.}, \bibinfo{author}{Leder, M.}, \bibinfo{author}{Grossert, C.},
 \bibinfo{author}{Weitz, M.} \& \bibinfo{author}{Solano, E.}
\newblock \bibinfo{title}{Quantum Rabi model in the Brillouin zone with ultracold atoms}.
\newblock \emph{\bibinfo{journal}{Phys. Rev. A}}
  \textbf{\bibinfo{volume}{95}}, \bibinfo{pages}{013827}
  (\bibinfo{year}{2017}).

\bibitem{Schneeweiss2018}
\bibinfo{author}{Schneeweiss, P.}, \bibinfo{author}{Dareau, A.} \& \bibinfo{author}{Sayrin, C.}
\newblock \bibinfo{title}{Cold-atom based implementation of the quantum Rabi model}.
\newblock \emph{\bibinfo{journal}{Phys. Rev. A}}
  \textbf{\bibinfo{volume}{98}}, \bibinfo{pages}{021801}
  (\bibinfo{year}{2018}).

\bibitem{Leggett2002}
\bibinfo{author}{Leggett, A.-J.}
\newblock \bibinfo{title}{TOPICAL REVIEW: Testing the limits of quantum mechanics: motivation, state of play, prospects}.
\newblock \emph{\bibinfo{journal}{Journal of Physics Condensed Matter}}
  \textbf{\bibinfo{volume}{14}}, \bibinfo{pages}{R415-R451}
  (\bibinfo{year}{2002}).

\bibitem{Armour2002}
\bibinfo{author}{Armour, A.-D.}, \bibinfo{author}{Blencowe, M.-P.} \& \bibinfo{author}{Schwab, K.-C.}
\newblock \bibinfo{title}{Entanglement and Decoherence of a Micromechanical Resonator via Coupling to a Cooper-Pair Box}.
\newblock \emph{\bibinfo{journal}{Phys. Rev. Lett.}}
  \textbf{\bibinfo{volume}{88}}, \bibinfo{pages}{148301}
  (\bibinfo{year}{2002}).

\bibitem{Liao2016}
\bibinfo{author}{Liao, J.-Q.}, \bibinfo{author}{Huang, J.-F.} \& \bibinfo{author}{Tian, L.}
\newblock \bibinfo{title}{Generation of macroscopic Schr\"odinger-cat states in qubit-oscillator systems}.
\newblock \emph{\bibinfo{journal}{Phys. Rev. A}}
  \textbf{\bibinfo{volume}{93}}, \bibinfo{pages}{033853}
  (\bibinfo{year}{2016}).

\bibitem{Haljan2005}
\bibinfo{author}{Haljan, P.-C.}, \bibinfo{author}{Brickman, K.-A.}, \bibinfo{author}{Deslauriers, L.}, \bibinfo{author}{Lee, P.-J.} \& \bibinfo{author}{Monroe, C.}
\newblock \bibinfo{title}{Spin-Dependent Forces on Trapped Ions for Phase-Stable Quantum Gates and Entangled States of Spin and Motion}.
\newblock \emph{\bibinfo{journal}{Phys. Rev. Lett.}}
  \textbf{\bibinfo{volume}{94}}, \bibinfo{pages}{153602}
  (\bibinfo{year}{2005}).

\bibitem{Yin2013}
\bibinfo{author}{Yin, Z.-q.}, \bibinfo{author}{Li, T.}, \bibinfo{author}{Zhang, X.} \& \bibinfo{author}{Duan, L.-M.}
\newblock \bibinfo{title}{Large quantum superpositions of a levitated nanodiamond through spin-optomechanical coupling}.
\newblock \emph{\bibinfo{journal}{Phys. Rev. A}}
  \textbf{\bibinfo{volume}{88}}, \bibinfo{pages}{033614}
  (\bibinfo{year}{2013}).

\bibitem{Liu2005}
\bibinfo{author}{Liu, Y.-X.}, \bibinfo{author}{Wei, L.-F.} \& \bibinfo{author}{Nori, F.}
\newblock \bibinfo{title}{Preparation of macroscopic quantum superposition states of a cavity field via coupling to a superconducting charge qubit}.
\newblock \emph{\bibinfo{journal}{Phys. Rev. A}}
  \textbf{\bibinfo{volume}{71}}, \bibinfo{pages}{063820}
  (\bibinfo{year}{2005}).

\bibitem{Liao2008}
\bibinfo{author}{Liao, J.-Q.} \& \bibinfo{author}{Kuang, L.-M.}
\newblock \bibinfo{title}{Nanomechanical resonator coupling with a double quantum dot: quantum state engineering}.
\newblock \emph{\bibinfo{journal}{The European Physical Journal B}}
  \textbf{\bibinfo{volume}{63}}, \bibinfo{pages}{79}
  (\bibinfo{year}{2008}).

\bibitem{Sorensen2000}
\bibinfo{author}{S{\o}rensen, A.} \& \bibinfo{author}{M{\o}lmer, K.}
\newblock \bibinfo{title}{Entanglement and quantum computation with ions in thermal motion}.
\newblock \emph{\bibinfo{journal}{Phys. Rev. A}}
  \textbf{\bibinfo{volume}{62}}, \bibinfo{pages}{022311}
  (\bibinfo{year}{2000}).

\bibitem{Zoller2003}
\bibinfo{author}{Garc\'{\i}a-Ripoll, J.-J.}, \bibinfo{author}{Zoller, P.} \& \bibinfo{author}{Cirac, J.-I.}
\newblock \bibinfo{title}{Speed Optimized Two-Qubit Gates with Laser Coherent Control Techniques for Ion Trap Quantum Computing}.
\newblock \emph{\bibinfo{journal}{Phys. Rev. Lett.}}
  \textbf{\bibinfo{volume}{91}}, \bibinfo{pages}{157901}
  (\bibinfo{year}{2003}).

\bibitem{Leibfried2003}
\bibinfo{author}{Leibfried, D.}, \bibinfo{author}{DeMarco, B.}, \bibinfo{author}{Meyer, V.}, \bibinfo{author}{Lucas, D.}, \bibinfo{author}{Barrett, M.}, \bibinfo{author}{Britton, J.}, \bibinfo{author}{Itano, W.-M.}, \bibinfo{author}{Jelenkovi\'{c}, B.}, \bibinfo{author}{Langer, C.}, \bibinfo{author}{Rosenband, T.} \& \bibinfo{author}{Wineland, D.-J.}
\newblock \bibinfo{title}{Experimental demonstration of a robust, high-fidelity geometric two ion-qubit phase gate}.
\newblock \emph{\bibinfo{journal}{Nature}}
  \textbf{\bibinfo{volume}{422}}, \bibinfo{pages}{412}
  (\bibinfo{year}{2003}).

\bibitem{Feng2007}
\bibinfo{author}{Feng, X.-L.}, \bibinfo{author}{Wang, Z.-S.}, \bibinfo{author}{Wu, C.-F.}, \bibinfo{author}{Kwek, L.-C.}, \bibinfo{author}{Lai, C.-H.} \& \bibinfo{author}{Oh, C.-H.}
\newblock \bibinfo{title}{Scheme for unconventional geometric quantum computation in cavity QED}.
\newblock \emph{\bibinfo{journal}{Phys. Rev. A}}
  \textbf{\bibinfo{volume}{75}}, \bibinfo{pages}{052312}
  (\bibinfo{year}{2007}).

\bibitem{Feng2009}
\bibinfo{author}{Feng, X.-L.}, \bibinfo{author}{Wu, C.}, \bibinfo{author}{Sun, H.} \& \bibinfo{author}{Oh, C.-H.}
\newblock \bibinfo{title}{Geometric Entangling Gates in Decoherence-Free Subspaces with Minimal Requirements}.
\newblock \emph{\bibinfo{journal}{Phys. Rev. Lett.}}
  \textbf{\bibinfo{volume}{103}}, \bibinfo{pages}{200501}
  (\bibinfo{year}{2009}).

\bibitem{Billangeon2015}
\bibinfo{author}{Billangeon, P.-M.}, \bibinfo{author}{Tsai, J.-S.} \& \bibinfo{author}{Nakamura, Y.}
\newblock \bibinfo{title}{Circuit-QED-based scalable architectures for quantum information processing with superconducting qubits}.
\newblock \emph{\bibinfo{journal}{Phys. Rev. B}}
  \textbf{\bibinfo{volume}{91}}, \bibinfo{pages}{094517}
  (\bibinfo{year}{2015}).

\bibitem{Zhu2003}
\bibinfo{author}{Zhu, S.-L.} \& \bibinfo{author}{Wang, Z.-D.}
\newblock \bibinfo{title}{Unconventional Geometric Quantum Computation}.
\newblock \emph{\bibinfo{journal}{Phys. Rev. Lett.}}
  \textbf{\bibinfo{volume}{91}}, \bibinfo{pages}{187902}
  (\bibinfo{year}{2003}).

\bibitem{Kirchmair2009}
\bibinfo{author}{Kirchmair, G.}, \bibinfo{author}{Benhelm, J.}, \bibinfo{author}{Z\"{a}hringer, F.}, \bibinfo{author}{Gerritsma, R.}, \bibinfo{author}{Roos, C.-F.} \& \bibinfo{author}{Blatt, R.}
\newblock \bibinfo{title}{Deterministic entanglement of ions in thermal states of motion}.
\newblock \emph{\bibinfo{journal}{New Journal of Physics}}
  \textbf{\bibinfo{volume}{11}}, \bibinfo{pages}{023002}
  (\bibinfo{year}{2009}).

\bibitem{Wang2010}
\bibinfo{author}{Wang, Y.-D.}, \bibinfo{author}{Chesi, S.}, \bibinfo{author}{Loss, D.} \& \bibinfo{author}{Bruder, C.}
\newblock \bibinfo{title}{One-step multiqubit Greenberger-Horne-Zeilinger state generation in a circuit QED system}.
\newblock \emph{\bibinfo{journal}{Phys. Rev. B}}
  \textbf{\bibinfo{volume}{81}}, \bibinfo{pages}{104524}
  (\bibinfo{year}{2010}).

\bibitem{MS1999}
\bibinfo{author}{M\o{}lmer, K.} \& \bibinfo{author}{S\o{}rensen, A}
\newblock \bibinfo{title}{Multiparticle Entanglement of Hot Trapped Ions}.
\newblock \emph{\bibinfo{journal}{Phys. Rev. Lett.}}
  \textbf{\bibinfo{volume}{82}}, \bibinfo{pages}{1835}
  (\bibinfo{year}{1999}).

\bibitem{Wang2002-1}
\bibinfo{author}{Wang, X.-G.} \& \bibinfo{author}{Zanardi, P.}
\newblock \bibinfo{title}{Simulation of many-body interactions by conditional geometric phases}.
\newblock \emph{\bibinfo{journal}{Phys. Rev. A}}
  \textbf{\bibinfo{volume}{65}}, \bibinfo{pages}{032327}
  (\bibinfo{year}{2002}).

\bibitem{Christian2008}
\bibinfo{author}{Christian, F.-R.}
\newblock \bibinfo{title}{Ion trap quantum gates with amplitude-modulated laser beams}.
\newblock \emph{\bibinfo{journal}{New Journal of Physics}}
  \textbf{\bibinfo{volume}{10}}, \bibinfo{pages}{013002}
  (\bibinfo{year}{2008}).

\bibitem{Blanes2009}
\bibinfo{author}{Blanes, S.}, \bibinfo{author}{Casas, F.}, \bibinfo{author}{Oteo, J.-A.} \& \bibinfo{author}{Ros, J.}
\newblock \bibinfo{title}{The Magnus expansion and some of its applications}.
\newblock \emph{\bibinfo{journal}{Physics Reports}}
  \textbf{\bibinfo{volume}{470}}, \bibinfo{pages}{151}
  (\bibinfo{year}{2009}).

\bibitem{Liu2007}
\bibinfo{author}{Liu, Y.}, \bibinfo{author}{Wei, L.}, \bibinfo{author}{Johansson, J.}, \bibinfo{author}{Tsai, J.} \& \bibinfo{author}{Nori, F.}
\newblock \bibinfo{title}{Superconducting qubits can be coupled and addressed as trapped ions}.
\newblock \emph{\bibinfo{journal}{Phys. Rev. B}}
  \textbf{\bibinfo{volume}{76}}, \bibinfo{pages}{144518}
  (\bibinfo{year}{2007}).

\bibitem{Goerz2017}
\bibinfo{author}{Goerz, M.}, \bibinfo{author}{Motzoi, F.}, \bibinfo{author}{Whaley, K.} \& \bibinfo{author}{Koch, C.} 
\newblock \bibinfo{title}{Charting the circuit QED design landscape using optimal control theory}.
\newblock \emph{\bibinfo{journal}{npj Quantum Information}}
  \textbf{\bibinfo{volume}{3}}, \bibinfo{pages}{37}
  (\bibinfo{year}{2017}).

\bibitem{Zanardi2000}
\bibinfo{author}{Zanardi, P.}, \bibinfo{author}{Zalka, C.} \& \bibinfo{author}{Faoro, L.}
\newblock \bibinfo{title}{Entangling power of quantum evolutions}.
\newblock \emph{\bibinfo{journal}{Phys. Rev. A}}
  \textbf{\bibinfo{volume}{62}}, \bibinfo{pages}{030301}
  (\bibinfo{year}{2000}).

\bibitem{Zanardi2001}
\bibinfo{author}{Zanardi, P.}
\newblock \bibinfo{title}{Entanglement of quantum evolutions}.
\newblock \emph{\bibinfo{journal}{Phys. Rev. A}}
  \textbf{\bibinfo{volume}{63}}, \bibinfo{pages}{040304}
  (\bibinfo{year}{2001}).

\bibitem{Wang2002-2}
\bibinfo{author}{Wang, X.} \& \bibinfo{author}{Zanardi, P.}
\newblock \bibinfo{title}{Quantum entanglement of unitary operators on bipartite systems}.
\newblock \emph{\bibinfo{journal}{Phys. Rev. A}}
  \textbf{\bibinfo{volume}{66}}, \bibinfo{pages}{044303}
  (\bibinfo{year}{2002}).

\bibitem{Ma2007}
\bibinfo{author}{Ma, Z.} \& \bibinfo{author}{Wang, X.}
\newblock \bibinfo{title}{Matrix realignment and partial-transpose approach to entangling power of quantum evolutions}.
\newblock \emph{\bibinfo{journal}{Phys. Rev. A}}
  \textbf{\bibinfo{volume}{75}}, \bibinfo{pages}{014304}
  (\bibinfo{year}{2007}).

\bibitem{Makhlin2002}
\bibinfo{author}{Makhlin, Y.}
\newblock \bibinfo{title}{Nonlocal Properties of Two-Qubit Gates and Mixed States, and the Optimization of Quantum Computations}.
\newblock \emph{\bibinfo{journal}{Quantum Information Processing}}
  \textbf{\bibinfo{volume}{1}}, \bibinfo{pages}{243}
  (\bibinfo{year}{2002}).

\bibitem{Zhang2003}
\bibinfo{author}{Zhang, J.}, \bibinfo{author}{Vala, J.}, \bibinfo{author}{Sastry, S.} \& \bibinfo{author}{Whaley, K.-B.}
\newblock \bibinfo{title}{Geometric theory of nonlocal two-qubit operations}.
\newblock \emph{\bibinfo{journal}{Phys. Rev. A}}
  \textbf{\bibinfo{volume}{67}}, \bibinfo{pages}{042313}
  (\bibinfo{year}{2003}).
\end{thebibliography}

\section*{Acknowledgements}
The work is supported by the NSF of China (Grant No. 11575042).

\section*{Author Contributions}
G.W., C.W., and K.X. initiated the idea. C.S., J.L. and R.X. developed the model and performed the calculations. J.L. and R.X. provided numerical results. All authors developed the scheme and wrote the main manuscript text.

\vspace{2pt}

\section*{Additional Information}
\textbf{Competing Interests:} The authors declare no competing interests.

\end{document}